\documentclass[prl,superscriptaddress,a4paper,twocolumn,showpacs ]{revtex4}
\usepackage{amsmath,amsfonts,bbm, graphicx,color}
\def\>{\rangle}
\def\<{\langle}

\def\A{ {\cal A} }
\def\Q{ {\cal Q} }
\def\P{ {\cal P} }

\def\T{ {\cal T} }

\def\Tr{ \mbox{Tr} }
\def\non{ \nonumber\\}

\begin{document}

\title{Entanglement and the Thermodynamic Arrow of Time.} %

\author{David Jennings}%
\affiliation{Institute for Mathematical Sciences, Imperial College London, London SW7 2BW,
United Kingdom}%
\author{Terry Rudolph}%
\affiliation{Institute for Mathematical Sciences, Imperial College London, London SW7 2BW,
United Kingdom}%

\date{January 2010}

\begin{abstract}
We discuss quantum entanglement in the context of the thermodynamic arrow of time. We review the role of correlations in entropy-decreasing events and prove that the occurrence of a transformation between two thermodynamic states constitutes a new type of entanglement witness, one not defined as a separating plane in state space between separable and entangled states, but as a physical process dependent on the local initial properties of the states. Extending work by Partovi, we consider a general entangled multipartite system that allows large reversals of the thermodynamic arrow of time. We describe a hierarchy of arrows that arises from the different correlations allowed in a quantum state and examine these features in the context of Maxwell's Demon. We examine in detail the case of three qubits, and also propose some simple experimental demonstrations possible with small numbers of qubits.
\end{abstract}

\maketitle

\section{Introduction}

For over a century, Maxwell's demon has provided a setting in which to address the limitations that thermodynamics places on an observer free to perform measurements on a system and then act on their acquired information in some algorithmic way \cite{maxwell-1867, maxwell-1871, szilard-1929, vedral-review}. As is now well appreciated, information has an energetic value. A demon \footnote{Appropriately enough, Plato argued that the word `demon' comes from the Greek word \textit{da$\bar{e}$mones} meaning \textit{`one who has knowledge or wisdom'}.} in possession of information about a physical state may transform this information into mechanical work \cite{szilard-1929, zurek-quantum-szilard}. Specifically, given a system of dimension $D$ in a state $\rho$ and the presence of a reservoir at a temperature $T$, the demon can extract $W=kT(\ln D - S[\rho])$ amount of mechanical work from the reservoir, where $S[\rho]=-\Tr [\rho \log \rho]$ is the von Neumann entropy of the state. The demon fails in his attempt to violate the second law of thermodynamics because the demon's memory, where he records the measurement results of the state, must be erased in order to operate in a cycle \cite{bennett82}. Such an erasure of memory can only occur with an unavoidable dissipation of heat \cite{landauer}.

As a concrete example, we may consider a gas of $N$ particles that resides in a piston on one side of a movable partition and in thermal equilibrium with a reservoir at a temperature $T$, Fig.~\ref{piston1}.
The demon, upon discovering what side of the partition the gas is on, may extract usable energy by deterministically acting on the piston system - if the gas is to the left of the partition, the demon puts the piston in contact with the reservoir, loads the partition with a mass $M$, tilts the piston as in Fig.~\ref{piston1} and allows the gas to expand slowly through the available empty volume. The expansion of the gas does work on the mass, elevating to a height $h$, which makes mechanical energy available. Indeed the expansion can provide a maximum mechanical energy of $NkT\ln 2$, which is extracted from the thermal reservoir.

\subsection{Confounding Maxwell's Demon}
In order to illustrate the new possibilities that quantum entanglement can bring to the story, it is amusing to consider cases in which a ``global demon'', with access to a large entangled quantum state, can confuse the traditional ``local demon'' who can only measure and act locally. The global demon can arrange that any thermodynamic process for the local demon can run ``backwards'' - heat can flow from a cold to hot and gases can contract instead of expand - driven by entanglement present in the state.

\begin{figure}[t]
\centering
\includegraphics[width=2cm]{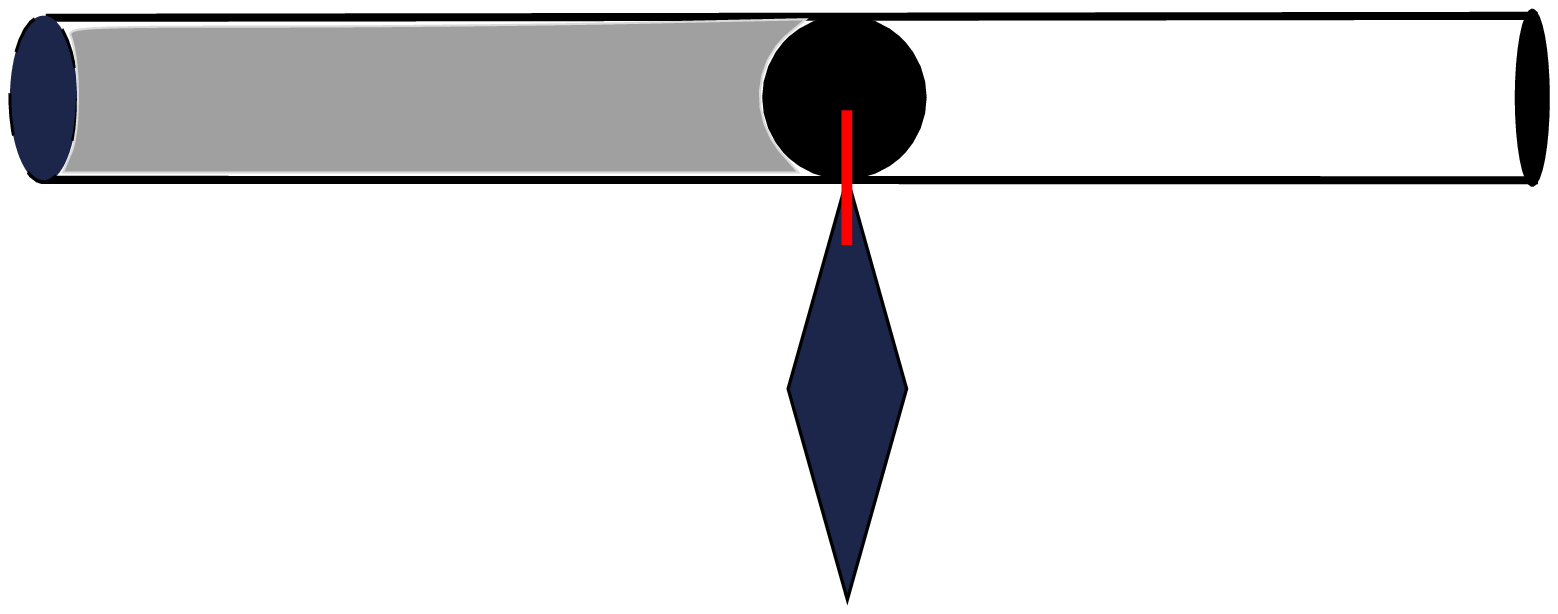}
\includegraphics[width=2cm]{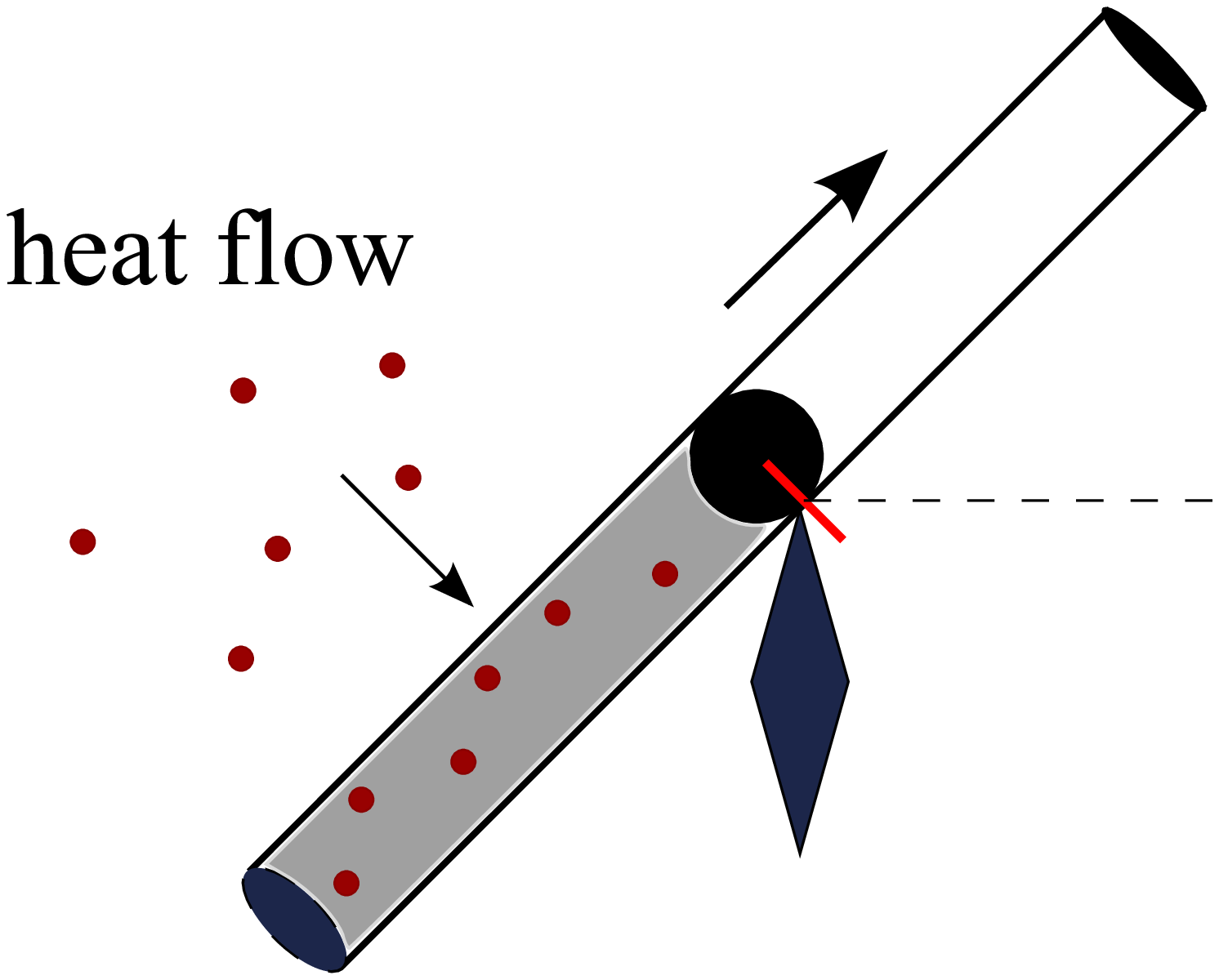}
\includegraphics[width=2cm]{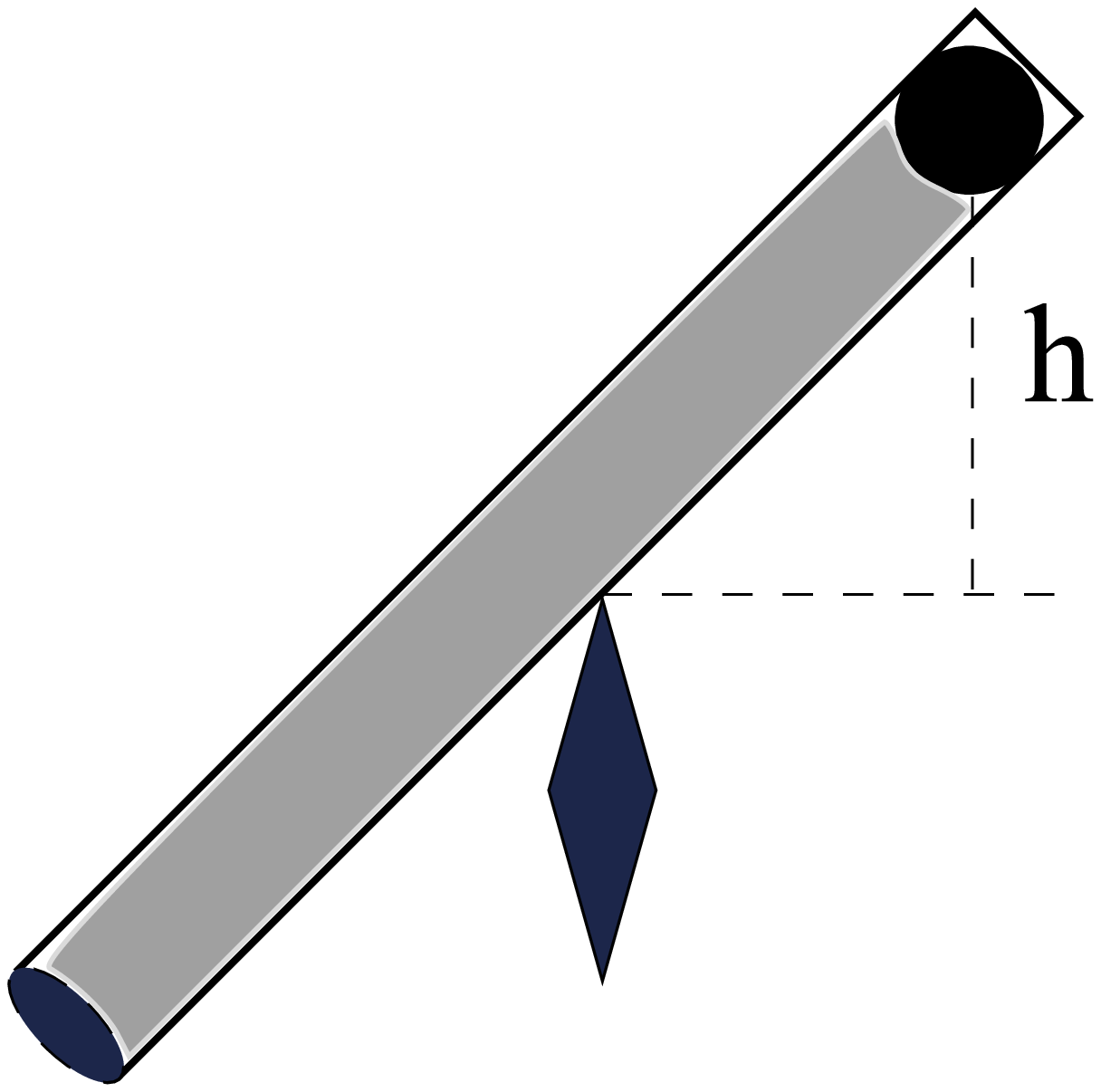}
\caption{\textbf{Uncorrelated case:} The piston is in contact with a thermal reservoir of particles, at temperature $T$. Mechanical energy is extracted from the reservoir by the demon choosing which way to tilt the piston, conditional on his information about what side the gas is located.}
\label{piston1}
\end{figure}

 As a concrete example, consider a system that consists of a reservoir and piston that are both in local thermal states, but with the global state of the system being pure. This state is prepared by the global demon, who may manipulate the global state as he pleases. The global demon ensures that the entanglement is local, in the sense that any part of the total system is entangled with its surroundings. Furthermore, for simplicity, we constrain the reservoir subsystems to couple to each other much weaker than to the piston subsystem \footnote{Although we shall see later that this requirement is not needed.}. This situation is indicated in Fig.~\ref{piston1}.

Since thermality arises from entanglement between the different components of the system in a global pure state, when the local demon places the piston in contact with the quantum reservoir and releases the partition, instead of the gas expanding up the piston to fill the empty volume, the thermal state repeatedly expands forwards and then contracts backwards in the tube as energy flows into and out of the piston. The spontaneous contraction phase of the gas is identical to the usual spontaneous expansion of a gas but with time ``running backwards''.

For the local demon, the thermodynamic arrow of time has failed and he is unable to extract any net work despite having been given a piston that is on its own completely indistinguishable from a standard thermal state on one side of a partition, seemingly coupled to a standard thermal reservoir.

The mischievous global demon, however, may further confuse the local demon by thermalizing the entanglement in the (quantum) reservoir that he has access to, using a different (fully thermal, classical) reservoir. Since the full system of dimension $D$ is in a pure state, the global demon can extract $kT \ln D $ amount of work from the classical reservoir, which he may store as mechanical energy, leaving the quantum reservoir plus piston in a maximally mixed state. Then using only some of this mechanical energy he may convert the maximally mixed reservoir into a true thermal state at the original temperature $T$. At which point the thermodynamic arrow is returned, the global demon has converted the entanglement into mechanical energy and the local demon, unaware of all the entanglement that was present, will finally be able to extract $NkT \ln 2$ of work from the reservoir as he had originally hoped. 

\begin{figure}[t]
\centering
\includegraphics[width=3cm]{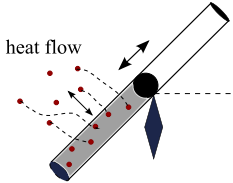}
\caption{\textbf{Entangled case:} The local thermodynamic states of the piston and reservoir are indistinguishable from the uncorrelated cases, however now entanglement is present in the system. In this case the piston both expands forwards and contracts backwards. Heat flows both into and out of the piston and the usual thermodynamic arrow no longer holds.}
\label{piston2}
\end{figure}

\subsection{The Arrow of Time}

There are many reasons to be interested in the arrow of time beyond simply fun stories about demons. The standard thermodynamic arrow applies to systems that are in isolation and amounts to a directionality for certain physical processes. The curious fact that certain processes are forbidden and certain states inaccessible is generally traced back to the initial conditions of the physical system. Furthermore, since all parts of the observed universe seem to obey a common ``arrow of time'', the problem of resolving the thermodynamic arrow's origin requires us to account for the particular initial conditions of our universe \cite{penrose-weyl-hyp}.

The notion of an ``arrow of time'' is quite broad - extending from the uniform large-scale expansion of the universe, to the radiative arrow, and through to the individual, psychological perception of time \cite{Zeh-book, Huw-price-book}. However, since the laws of physics are invariant under time reversal (strictly CPT invariant), it is believed that any form of temporal directionality will stem ultimately from the initial conditions of the universe.

Recent work by Partovi \cite{partovi} has considered a specific pure entangled state that produced heat flow from a cold body to a hotter one, contrary to the usual thermodynamic arrow. In \cite{partovi} Partovi identifies the large degree of entanglement present in the state as being responsible for this unusual behaviour. We will see, in fact, that the negative heat flow he considers does not necessarily depend on the presence of entanglement in the state and, by adopting an information theoretic framework, we explore these and similar possibilities as a way of elucidating the properties of entangled quantum states and the connection with varied notions of ``arrows of time''.

\subsection{The Thermodynamic Arrow and Correlations}

A paradigmatic setting that captures the essentials of the thermodynamic arrow of time is in the context of heat flow between two systems $A$ and $B$ that interact in isolation, so that their total energy is constant. For simplicity, we assume that they do no mechanical work on each other, but may exchange energy in the form of heat, for which conservation of energy implies that $Q_A+Q_B =0$, where $Q_A$ is the heat acquired by $A$ and $Q_B$ is the heat acquired by $B$.

Ultimately, the key ingredient that dictates which states are accessible via the set of all interactions on the total state $\rho_{AB}$ is the degree of correlations initially present between $A$ and $B$. These correlations between the two subsystems are quantified by the \textit{mutual information}, which is non-negative for all states and defined in terms of relative entropy as $I(A:B) = S[\rho_{AB} || \rho_A \otimes \rho_B]=S_A+S_B -S_{AB}$, where, for example, $S_A=-\Tr[\rho_A \log \rho_A]$ is the von Neumann entropy for the state $\rho_A$. The mutual information is a convenient measure of how distinguishable the state $\rho_{AB}$ is from the completely uncorrelated product state $\rho_A \otimes \rho_B$.

Of particular interest is when the reduced states at the initial time $t_i$ of subsystems $A$ and $B$ are thermal states for their respective individual Hamiltonians, $H_A$ and $H_B$. Given temperatures $T_A$ and $T_B$, the reduced states then take the form $\rho_A (t_i)\propto \exp [-\beta_A H_A]$ and $\rho_B (t_i) \propto \exp [-\beta_B H_B]$, where $\beta_X =1/kT_X$ for subsystem $X\in\{A,B\}$.

We may consider the most general process of switching on time-dependent interactions $V_{AB}(t)$ between $A$ and $B$ that evolve the composite system into a new state $\rho_{AB} (t)$.  For subsystem $X\in\{A,B\}$ at a later time $t$, the two thermodynamic variables of central interest are the average energy $U_X=\Tr[H_X \rho_X(t)]$ and the entropy $S_X=-\Tr[\rho_X(t) \log \rho_X(t)]$. Locally at $A$ the interaction results in the transformation $\rho_A(t_i) \rightarrow \rho_A(t_f) = \Tr_B [\rho_{AB}(t_f)]$, and so $U_A \rightarrow U_A +\Delta U_A$ and $S_A \rightarrow S_A +\Delta S_A$, with similar expressions for $B$.

These changes in local thermodynamic variables are constrained by the fact that a thermal state $\exp [- \beta H]/\Tr[\exp[-\beta H]]$ minimizes the free energy function $F[\rho] = \Tr[ \rho H] - S[\rho]/\beta$. Consequently we have that $\Tr[ \rho(t_i) H] - S[\rho(t_i)]/\beta<\Tr[ \rho(t_f) H] - S[\rho(t_f)]/\beta$, which more explicitly yields
\begin{eqnarray}\label{freee}
\beta_A \Delta U_A - \Delta S_A &\ge& 0 \nonumber \\
\beta_B \Delta U_B - \Delta S_B &\ge& 0
\end{eqnarray}
for the change in the thermodynamic variables defined at $A$ and $B$. It must be emphasized that equations (\ref{freee}) are kinematical restrictions on \emph{any} transformation from initially thermal states, $\{\rho_A (t_i),\rho_B (t_i)$\},  to any other pair of local states $\{\rho_A(t_f),\rho_B(t_f)\}$.

We consider transformations on the composite system that satisfy $S[\rho_{AB}(t_i)] =S[\rho_{AB}(t_f)]$ (which includes unitary transformations) and so $\Delta I(A:B) = \Delta S_A +\Delta S_B$. The only other restriction that we place on the transformation is that $\Tr[ \rho_{AB} (t_i) (H_A+H_B)]=\Tr[ \rho_{AB} (t_f) (H_A+H_B)]$. Since we don't consider mechanical work, this amounts to requiring that there is zero net heat flow into the composite system, and the total energy before the interaction process equals the total energy afterwards. Writing $Q_A= \Delta U_A$ and $Q_B= \Delta U_B$ for the heat gained by $A$ and $B$ respectively, we obtain for these energy preserving processes that
\begin{eqnarray}\label{flow}
\beta_A Q_A +\beta_B Q_B \ge \Delta I(A:B)
\end{eqnarray}
where $\Delta I(A:B)$ is the change in the mutual information between $A$ and $B$.

The special initial state $\rho_{AB}(t_i) = \rho_A(t_i) \otimes \rho_B(t_i)$ in which correlations vanish was historically called the condition of \textit{Stosszahlansatz}, or `molecular chaos'. If we consider $A$ and $B$ initially uncorrelated at some time $t_i$ then $I(A:B;t_i)=0$, however interactions allow correlations to form between the two subsystems, and so at later times $t_f$ we have $I(A:B;t_f) \ge 0$. We immediately deduce that $\beta_A Q_A +\beta_B Q_B \ge 0$, or more explicitly in terms of temperature
\begin{eqnarray}\label{uncorrelated}
Q_A \left ( \frac{1}{T_A} - \frac{1}{T_B}\right ) \ge 0.
\end{eqnarray}
In other words, heat can only flow from hot thermal states to cold thermal states in isolation. This is the standard thermodynamic arrow of time.

This situation can be generalized to $N$ initially uncorrelated thermal states and provides us with the constraint
\begin{eqnarray}\label{multiheat}
\sum _j  \frac{Q_j}{T_j} \ge 0
\end{eqnarray}
which forbids certain types of heat flow; for example, if we partition the $N$ thermal states in two, such that all the temperatures in the first group are lower than those in the second, then it is impossible for heat to flow from the first to the second.

In general, however, initial correlations are expected to be present, $I(A:B;t_i) >0$, and so the change in mutual information can be negative. From (\ref{flow}) it is clear that there is no longer a constraint on the directionality of heat flow between $A$ and $B$ \cite{seth-loyd}. As we shall now see, the correlations that make up the total mutual information can arise from classical correlations, or from a combination of classical correlations and quantum entanglement.

\subsection{Entanglement correlations}\label{entanglementcorrelations}

Entangled quantum systems can possess far stronger correlations than are possible classically, and in fact the vast majority of quantum states are entangled \cite{Clifton-separablevol, peres-book}. Work has been done previously on how quantum correlations affect macroscopic thermodynamic observables such as the susceptibility or the heat capacity, and it has been shown that measurement of these macroscopic observables can act as entanglement witnesses \cite{witness-1, witness-2}. We shall show that the thermodynamic \emph{transformations} themselves also constitute entanglement witnesses. We should be careful to distinguish this research program from the one to do with with formal analogies \cite{HOH, vedral02} between irreversible entanglement transformations and the second law of thermodynamics, on which much progress has recently been made \cite{fernando-entanglement-theory}, or from attempts to connect the second law of thermodynamics to Bell inequalities \cite{durham08}. Here we are interested in the physical effects which entanglement between systems and reservoirs has on the thermodynamical transformations of the systems. 

Recently in \cite{partovi} Partovi considered how the presence of entanglement in a system affects irreversible thermodynamic transformations.
The intriguing scenario he considered involves two subsystems $A$ and $B$ that are overall in a pure state $| \Psi _{AB} \>$ and possess strong quantum correlations. Furthermore, he arranged that the local reduced states $\rho_{A,B} = \Tr_{B,A} [|\Psi_{AB} \> \< \Psi_{AB} |]$ are thermal states for the two subsystems and so locally indistinguishable from classical thermal mixtures. Despite being locally indistinguishable from classical thermal mixtures, the composite system behaves quite differently under any energy-conserving unitary. Since the total initial state is a pure state we have that $S[\rho_{AB}(t_i)]=0$, which remains true at any later time. From a Schmidt decomposition for $A$ and $B$ we find that $\rho_A(t)$ and $\rho_B(t)$ are isospectral, $\mbox{Spec}(\rho_A(t))=\mbox{Spec}(\rho_B(t))$, and so their entropies are always equal. Consequently, any unitary transformation occurring on the composite system will have $\Delta S_A = \Delta S_B$.

Once again, assuming (i) no mechanical work and (ii) no overall change in the total energy, the condition on the heat flow between $A$ and $B$ is
 $\beta_A Q_A +\beta_B Q_B \ge \Delta S_A+\Delta S_B$.
However, since $\Delta S_A = \Delta S_B$ and $Q_A=-Q_B$ we have that
\begin{eqnarray}
-\Delta S_A( 1/\beta_A + 1/\beta_B)\ge 0
\end{eqnarray}
and so $\Delta S_A=\Delta S_B \le 0$. Energy conserving unitary interactions never \emph{increase} the local entropies of either subsystem, in stark contrast to the uncorrelated situation of classically mixed thermal states, where $\Delta S_A +\Delta S_B$ never decreases.

However, the model in \cite{partovi} does not actually require quantum entanglement. For example we may simply dephase $| \Psi_{AB} \>$ to turn it into a classically correlated separable state that would produce the exact same reversal of the thermodynamic arrow. In addition, while the pure state $| \Psi _{AB} \>$ allows local entropy decreases, and the reversal of the thermodynamic arrow, the isospectral constraint for the bipartite splitting of the pure state $| \Psi_{AB}\>$ is very restrictive - the only way in which two different temperatures for $A$ and $B$ is achieved in \cite{partovi} is to impose that the energy spectrum of $A$ be a scaled version of the energy spectrum of $B$. Another undesirable feature of the model is that the initial thermal states that are considered in \cite{partovi} (squeezed Gaussian states of two oscillators) are the \emph{only} thermal states that are attainable in this setting - all other thermal states are inaccessible.

In the next section we extend the work of \cite{partovi} to show that we may distinguish entangled systems from classically correlated system through violations of thermodynamic arrow and produce transformations between arbitrary thermal states without restrictions on the energy spectra. Furthermore, we analyze quantum systems in which the correlation structure is more complex and allows for a more subtle range of violations of the thermodynamic arrow.

\subsection{Classical correlations and entanglement witnesses}

Recently \cite{grois05} the mutual information $I(A:B)$ was given an operational meaning in terms of the minimal amount of local randomizing work that must be done on a bipartite system to reduce it to a product state and destroy all correlations. We will give here another thermodynamical interpretation - namely, the mutual information is a measure of the maximal amount that the isolated bipartite system may violate the thermodynamic arrow in the form of energy transfer from a colder subsystem to a hotter one.

The total correlations, as quantified by $I(A:B)$, also contain classical correlations, which may be defined as
\begin{eqnarray}
I_c(A:B) = \mbox{max}_{M_A\otimes M_B} H(A:B)
\end{eqnarray}
where $H(A:B)$ is the classical (Shannon) mutual information for the joint probability distributions generated by local POVM measurements $M_A \otimes M_B$ at $A$ and $B$ \cite{definition-of-Ic}. The classical mutual information $I_c(A:B)$ is always less than or equal to the total mutual information $I(A:B)$ and vanishes if and only if the bipartite state is a product state\cite{locking-classical}.

It is clear that for bipartite states $\rho_{AB}$ the classical mutual information always obeys the bound $ 0\le I_c(A:B) \le \log D$, where $D$ is the dimensionality of the smaller subsystem. On the other hand, while the quantum mutual information $I(A:B)$ is always positive, it can in general take on values larger than $\log D$.  However, if we write the quantum mutual information as
\begin{eqnarray}
I(A:B) &=& S_A+S_B -S_{AB} \nonumber \\
&=& S_A - (S_{AB}-S_B) =S_B - (S_{AB}-S_A)\nonumber
\end{eqnarray}
and restrict ourselves to separable states $\rho_{AB}$ for which $S_{AB}$ is always greater than both $S_B$ and $S_A$ \cite{positive-conditional} we may deduce that $I(A:B) \le \mbox{min} \{S_A,S_B\}$ and so $I(A:B)|_{\mbox{\tiny sep}} \le \log D$ over the set of separable states. This upper bound is saturated for perfectly correlated, zero discord, separable states $\rho_{AB} = \frac{1}{D}\sum_i |i\>_A\<i| \otimes |i\>_B\<i|$, in which case $I_c(A:B) = I(A:B) = \log D$.

However, for entangled states we can have $I(A:B) > \log D$, and so from (\ref{flow}) we see that certain types of heat flow are possible that are forbidden classically. In particular a quantity of heat $I(A:B)/(|\beta_A -\beta_B|)$ can be transferred from the cold subsystem to the hot system, whereas classical correlations could only permit at most a quantity of heat $\log D/ (| \beta_A - \beta_B|)$ to flow from cold to hot. From this it follows that the reversal of the thermodynamic arrow of time is a new kind of entanglement witness where quantum correlations present in the state allow much larger violations of the arrow than are classically possible, and so the detection of such large reversals must imply an entangled state. Standard entanglement witnesses correspond to an observable $W$ that defines a plane $\Tr [W \rho] =0$ in state space separating entangled states from separable ones, while for us, it is the set of energetically accessible states that acts as the witness. Instead of local observables, it is the transformations of the state that signal entanglement.

The classical mutual information $I_c(A:B)$ has previously been considered in the context of the thermodynamic arrow as a candidate measure for the classical memory record that a system $A$ has of events that affect $B$ \cite{macc08}. The claim in \cite{macc08} was that any event that decreases the entropy of $B$ necessarily reduces the classical mutual information between $A$ and $B$, and so entropy-decreasing events at $B$ do not leave a memory trace in $A$. This was a proposed resolution to the empirical fact that we only observe entropy-decreasing events, namely the ``arrow of time dilemma''.

However, in \cite{jennings09} we demonstrate explicitly that this claim is false, and in reality quantum mechanics allows the exact opposite to occur. We show that large quantum correlations present in a state can be used up to reverse the thermodynamic arrow and at the same time increase the classical correlations between two subsystems. Quantum mechanics allows enhanced memory records of entropy-increasing events and so there can be no resolution of the arrow of time in terms of classical memory records.

The process of dividing correlations into quantum and classical components has many subtle aspects and counter-intuitive results exist, for example in the setting of multipartite correlations it is possible to have purely quantum multipartite correlations without any classical multipartite correlations being present \cite{ved03}. In the next section we consider multipartite quantum systems that are in a highly entangled pure state and analyse how the quantum correlations present allow arbitrarily large reversals of the thermodynamic arrow.

\section{Local Thermal States, Global States and a Hierarchy of Arrows}

 We now turn to an analysis of entanglement-assisted violations of the thermodynamic arrow in a general multipartite setting. Once again, we consider an isolated system that may undergo a complex time-dependent interaction process between times $t_i$ and $t_f$, but with the total energy, defined via a total Hamiltonian $H_{\mbox{\tiny tot}}$, at time $t_i$ equal to the total energy at time $t_f$.

It is well known that most quantum states are highly entangled \cite{Clifton-separablevol}, however recently a deep theorem \cite{pop03} has shown that if the total system is large, then any randomly choosen pure state $|\Phi\>$ that satisfies the total energy constraint has the property that any sufficiently small subsystem is highly likely to be in a thermal state. For generic large systems in a pure state, thermality of its subsystems naturally arises from the high degree of entanglement present in the state. Consequently, instead of entangled thermal states such as in \cite{partovi} being artificial or exceptional, for large systems in a randomly chosen pure state they are actually quite typical.

\subsection{Local thermal states from an entangled state}

For simplicity we shall consider the multiple subsystems $S_1, \dots S_N$, to be qubits, but it is clear that similar arguments apply to any higher dimensional subsystems.

We take the reduced states, obtained by tracing out the other subsystems, to be initially thermal and the total system is assumed to be in a pure state $|\Psi \>$ with a fixed energy. We refer to the reduced states $\rho_1, \rho_2, \dots$ as the \emph{thermal marginals} for the subsystems.

However, not all sets of thermal marginals may be obtained from a global pure state $|\Psi \>$. For the case of qubits, if the marginal states for the individual qubits are given by $\{ \rho_i \}$, then the necessary and sufficient conditions for the existence of a pure state $| \Psi \>$ such that $\Tr _{j \ne i} | \Psi \> \< \Psi |= \rho_i$ are in terms of the smallest eigenvalue of each subsystem, $\lambda_i = \mbox{min(Spec(}\rho_i))$. The conditions are \cite{higuchi03}
\begin{eqnarray}\label{constraints}
\lambda_i &\le&  \sum_{j \ne i} \lambda_j  \non
0\le \lambda_i &\le & 1/2 \hspace{0.5cm} i=1, \dots , N.
\end{eqnarray}

Without loss of generality, we take the Hamiltonians of each qubit to be equal, and choose the ground state to have zero energy while the excited state has energy 1. Thus, $H_i = \frac{1}{2} (I + Z_i)$, where $Z_i$ is the Pauli operator for the $Z$ direction,  and for qubits with thermal marginals, the total expected energy of the system is then $E =\sum _i \lambda_i $. The parameter space for the set of states $|\Psi \>$ in terms of the smallest eigenvalues of its subsystems is defined by (\ref{constraints}) and is an $N$-dimensional polyhedron $\P_N$, with the constant energy condition being a hyperplane that intersects $\P_N$ on a subset $\T_{N-1}$ of dimension at most $N-1$. Each point in $\T_{N-1}$ corresponds to an accessible combination of thermal marginals, however there is in general more than one pure global state associated to such a point.

\begin{figure}[t]
\centering
\includegraphics[width=5.5cm]{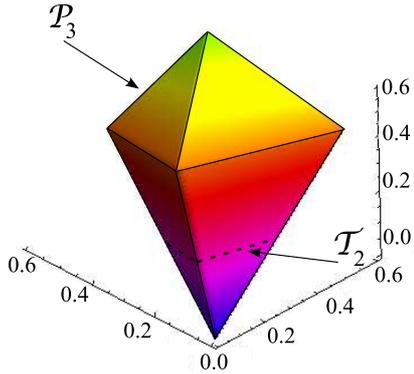}
\caption{\textbf{The parameter space $\P_3$:} The set of entangled states giving local thermal states possesses a diamond polygon parameter space. A triangular slice $\T_2$ with constant total energy is shown. Dark regions have low energies and temperatures, while bright regions have high energies and temperatures.}
\label{colorpoly}
\end{figure}

As an example, in the case $N=3$, the region of parameter space is a diamond formed from two tetrahedra, while the constant energy condition corresponds to a plane that slices the diamond in triangular cross-sections, as depicted in Fig.~\ref{colorpoly}.
Each point in the diamond defines three qubit thermal states, however, since the overall state is a pure state, any other pure state may be reached by a general unitary transformation. It is the restriction to energy-conserving unitaries which means that only thermal states with parameters in the triangular region $\T_2$ are accessible. Such energy conserving unitaries could arise from time-dependent interactions $V_{ij}(t)$, or, for example, from time-independent interactions of the form $X_i Y_j -Y_iX_j$.

The diamond polyhedron has some nice features. The origin is the bottom vertex and corresponds to $E=0$, where all the qubits are at $T=0$, while $E=3/2$ is the other extreme point, where each qubit is maximally mixed, and $T= \infty$.
The centroid of any given constant energy triangle corresponds to all three qubits having equal temperatures and so would correspond to the standard equilibrium one would expect if the three subsystems were initially uncorrelated. For standard thermally mixed states free to interact, the total system would tend to evolve towards this configuration. However, from our analysis we see that this is no longer imposed. The overall state is pure and the total system can undergo unitary evolution to any other set of local thermal states in $\T_2$. The large degree of entanglement present has radically lifted the constraints of the thermodynamic arrow and allows the subsystems to transformation to otherwise inaccessible states.

The triangle formed by the intersection of $E=1$ and the diamond forms the widest part of the polyhedron and is special in the sense that there is enough entanglement for two of the qubits to form a singlet state, but not too much that the singlet becomes impossible. In terms of temperature, for points on this triangle it is possible to unitarily transform along $E=$constant to a situation where two qubits are maximally mixed while the third is in a pure state. For $E<1$ it is always possible to turn off the temperature of one qubit with the other two being at some finite temperatures, while for $E>1$ this is no longer possible and each qubit always has some non-zero temperature.

\begin{figure}[t]
\centering
\includegraphics[width=7cm]{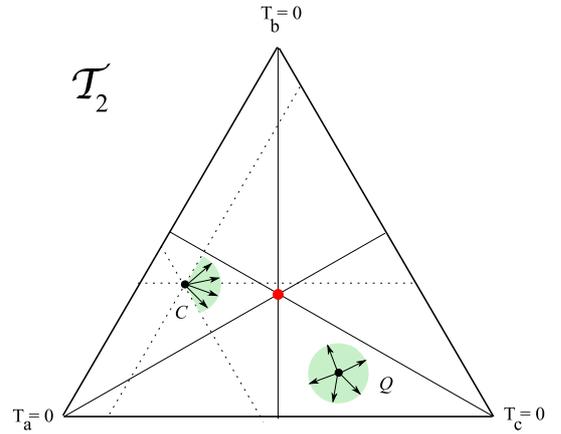}
\caption{\textbf{The parameter space $\T_2$:} The standard thermodynamic arrow would constrain $C$ to evolve only into the cone of states depicted, and this drives it steadily towards equilibrium, while the entangled state $Q$ has no such restriction. Dotted lines are the isotherms for $T_a$, $T_b$ and $T_c$, while the solid internal lines represent $(T_a=T_b,T_c=0)$, $(T_b=T_c,T_a=0)$ and $(T_c=T_a,T_b=0)$. The center point corresponds to thermal equilibrium $T_a=T_b=T_c$. }
\label{triarrows}
\end{figure}

\begin{figure}[t]
\centering
\includegraphics[width=6cm]{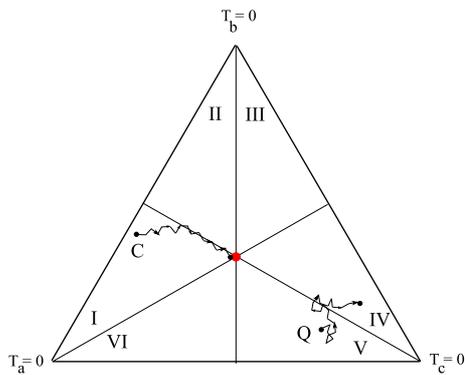}
\caption{Contrast in system evolution due to random transfers of energy between subsystems. For $C$, subject to the constraint (\ref{multiheat}), the thermodynamic arrow forces the subsystems to the equilibrium configuration. The system $Q$ is not restricted for the small fluctuations and performs a non-directed random walk through the accessible configuration space. }
\label{randomwalk}
\end{figure}

The diamond setting also provides a simple way to visualise the contrast between a system subject to the standard thermodynamic arrow and the highly entangled system discussed here. Fig.~\ref{triarrows} shows the constant energy region $\T_2$ for a system of three qubits with thermal marginals $\rho_a$, $\rho_b$ and $\rho_c$. The centroid of the triangle corresponds to the equilibrium configuration $T_a=T_b=T_c$. The solid lines passing through the equilibrium point and intersecting the midpoint of the sides correspond to the situation where two of the subsystems have equal temperatures. These lines divide the triangle into six regions, labelled I-VI in Fig.~\ref{randomwalk}, corresponding to the six choices of temperature orderings,  $T_a<T_b<T_c, \cdots$,  $T_c < T_b<T_a$.

If the system is in a thermal configuration $C$ with no correlations then equation (\ref{multiheat}) forces the system to evolve only in the direction shown in Fig.~\ref{triarrows}. The thermodynamic arrow therefore imposes a cone of accessible configurations for the thermal system, and gradually drives the system inexorably to thermal equilibrium at the center of the region (a fun comparison is with black holes - the light cone of a person inside the event horizon always forces them to move radially inwards, eventually reaching the central singularity, while here we have in some sense a `thermal cone' that dictates how the system must evolve in time, ultimately reaching the central equilibrium point).

It makes physical sense in this context to only consider a small transfer of energy that perturbs the initial thermal configuration into a neighbouring thermal configuration. A state with $T_b=T_c$ has no formal restriction from (\ref{multiheat}) on the amount of heat that may be exchanged between $b$ and $c$, however the configuration $T_b=T_c$ is stable in the sense that a small transfer of heat will automatically induce a correcting directionality that returns the system to $T_b=T_c$.

This is illustrated in Fig.~\ref{randomwalk} where given an initial thermal configuration $C$, subject to random heat exchanges and the constraint (\ref{multiheat}), the system starts in region I where $T_a < T_b < T_c$. Heat is exchanged, cooling $c$ and heating $a$ and $b$ until the system crosses the $T_b=T_c$ line into region II. At which point only fluctuations around this line are permitted by (\ref{multiheat}). Subsequently, the system $a$ is heated up until the thermal configuration reaches the central equilibrium point, as shown in the diagram, at which point only small fluctuations from equilibrium are permitted.

However, no constraint is present for our entangled state, which is free to execute a random walk and roam over the accessible region of parameter space $\T_2$, for example a given initial configuration $Q$ may start in region V, but is free to move towards and away from equilibrium and can visit every point in the accessible parameter space, see Fig.~\ref{randomwalk}.

\subsection{A global pure state}

We now present a highly entangled pure state for which the thermodynamic arrow is removed for interactions between any two of its subsystems, and so allows the most stark deviation from standard thermodynamical behaviour.

In general for $N$ qubits in locally thermal states there exist a vast number of consistent pure states. However, it is straightforward to see that any set of thermal marginals $\{ \rho_i \}_{i=1}^N $ can be obtained from a global pure state with a fixed energy $E$, of the form
\begin{eqnarray}\label{A1states}
| \Psi \> = \sum_i x_i X_i | 0 \cdots 0 \> + \sqrt{1 -E/(N-1)} |1 \cdots 1\>
\end{eqnarray}
with  $X_i$ the Pauli operator on qubit $i$, and each $x_i$ real and obeying the condition $\sum_i x_i^2 =E/(N-1)$.

Furthermore, these parameters are related to the parameters for $\P_N$ through $\lambda_i =\sum _{j\ne i} x^2_j$.

The parameters $x_i$ specify a point on a hypersphere in $\mathbb{R}^N$ of radius $\sqrt{E/(N-1)}$ and any unitary transformation that satisfies the energy constaint $\sum_i Q_i=0$ then corresponds to $O(N)$ transformations on the subspace spanned by $\{ X_i|0 \cdots 0\> \}_{i=1}^N$. Heat transfer between any two qubits $a$ and $b$ can be mediated by an energy-conserving interaction Hamiltonian such as $V_{ab}=(Y_a X_b - X_a Y_b)/2$ and the evolution is then described by the unitary $U_{ab} = \exp [ -i \theta/2 (Y_a X_b -X_a Y_b)]$. This unitary allows heat flow in either direction between the thermal reduced states $\rho_a$ and $\rho_b$.

While the reduced states $\rho_a, \rho_b$ are thermal, it is not the case that $\rho_{ab}= \Tr_{i \ne a,b} |\Psi \>\< \Psi|$ is a direct product of thermal states. The 2-qubit process we gave, detects the quantum correlations present between $a$ and $b$ that are not accounted for by merely probing the qubits individually. A state such as (\ref{A1states}) could clearly be used by a global demon to reverse standard thermodynamic behaviour for a local demon, as described in the introduction, with an interaction such as $V_{ab}$ causing heat to flow both into and out of a thermal state. In addition, we see that quantum correlations in a pure state allow far more freedom than would be allowed by classically correlated states.

\subsection{A candidate experimental state?}

\begin{figure}[t]
\centering
\includegraphics[width=4cm]{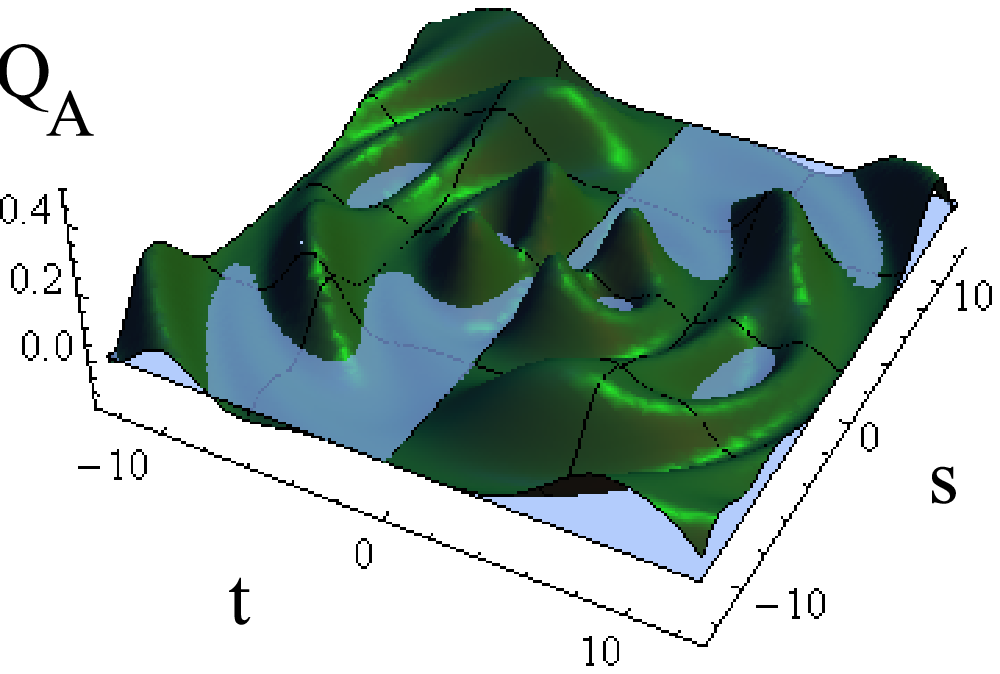}
\includegraphics[width=4cm]{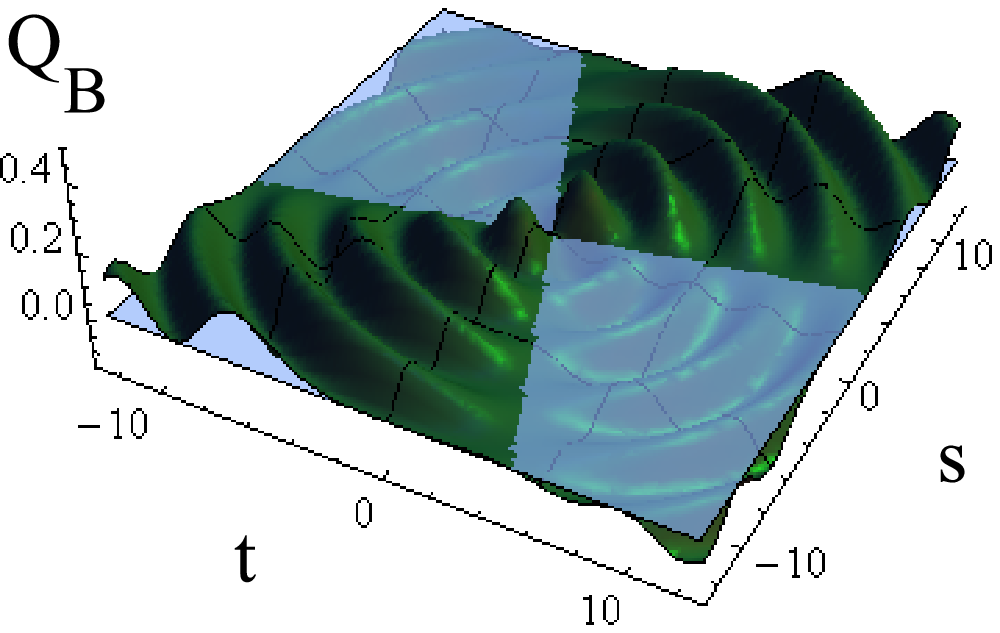}
\includegraphics[width=4cm]{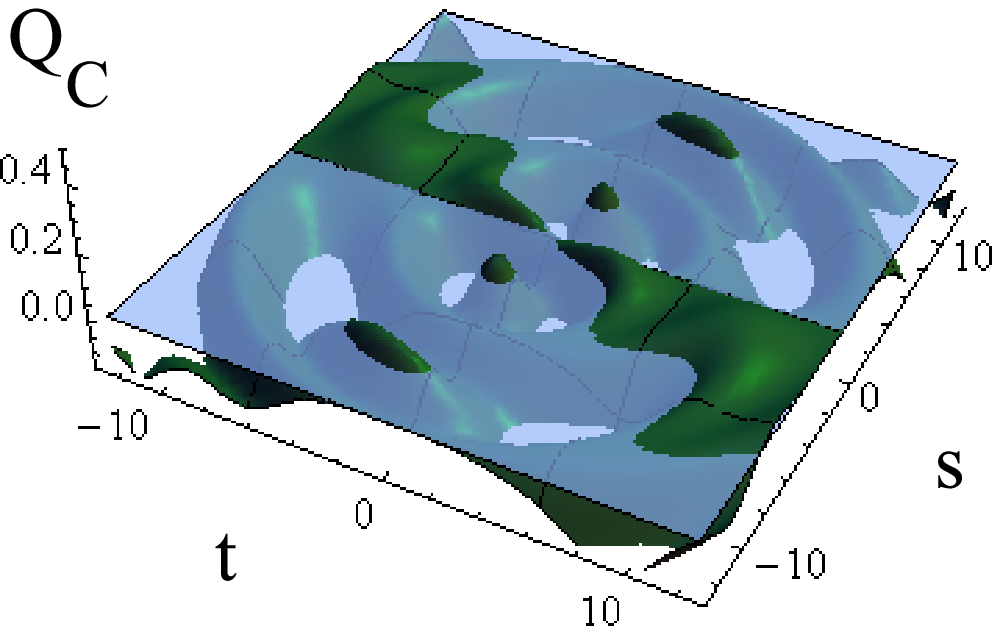}
\caption{\textbf{Extangled system:} The heat flow pattern into $A,B,C$ for the correlated state $\rho_{ABC}$ with $\lambda_A=0.15$, $\lambda_B=0.2 $, $\lambda_C=0.3$, $\gamma=0.4$ and parameters $t$ for the interaction $V_{AB}$ and $s$ for $V_{BC}$. Here a light-coloured plane is shown through $Q=0$ 
to highlight those regions where heat leaves the subsystem $Q<0$, while the dark regions visible above this plane correspond to heat entering the system $Q>0$.}
\label{correlatedABC}
\end{figure}

\begin{figure}[t]
\centering
\includegraphics[width=4cm]{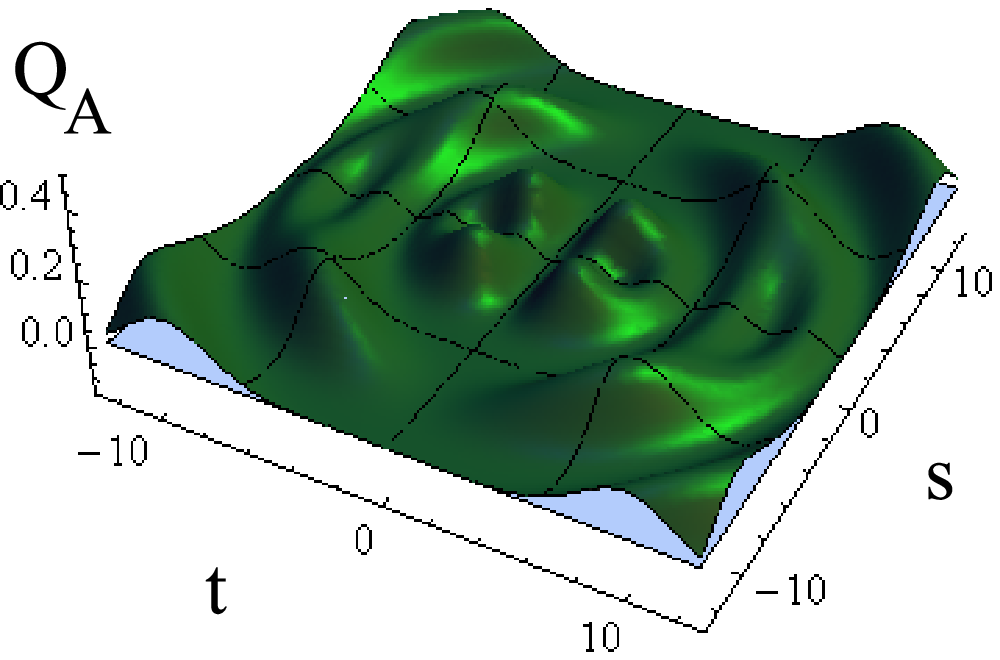}
\includegraphics[width=4cm]{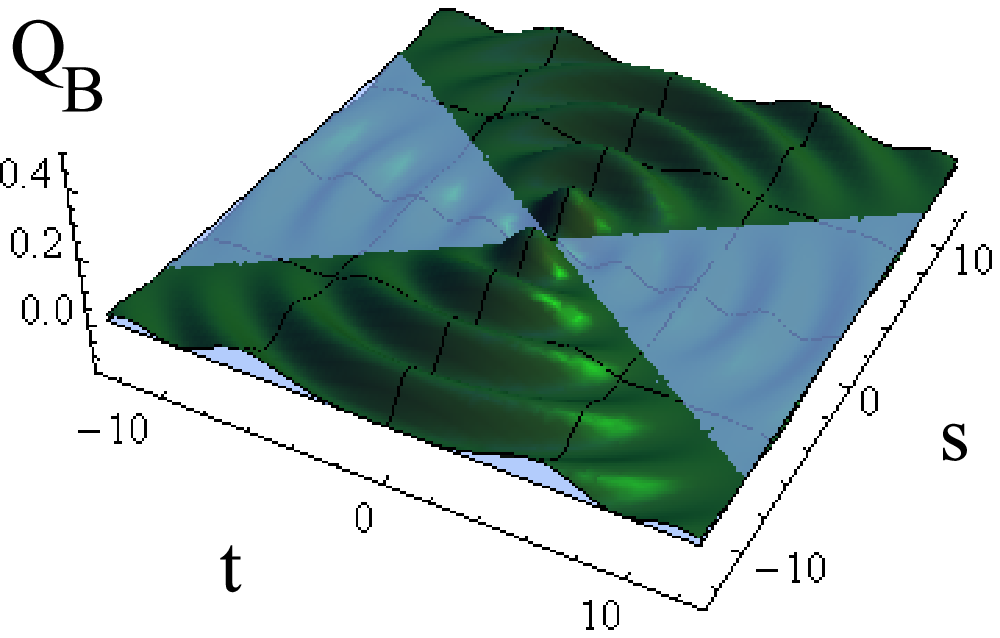}
\includegraphics[width=4cm]{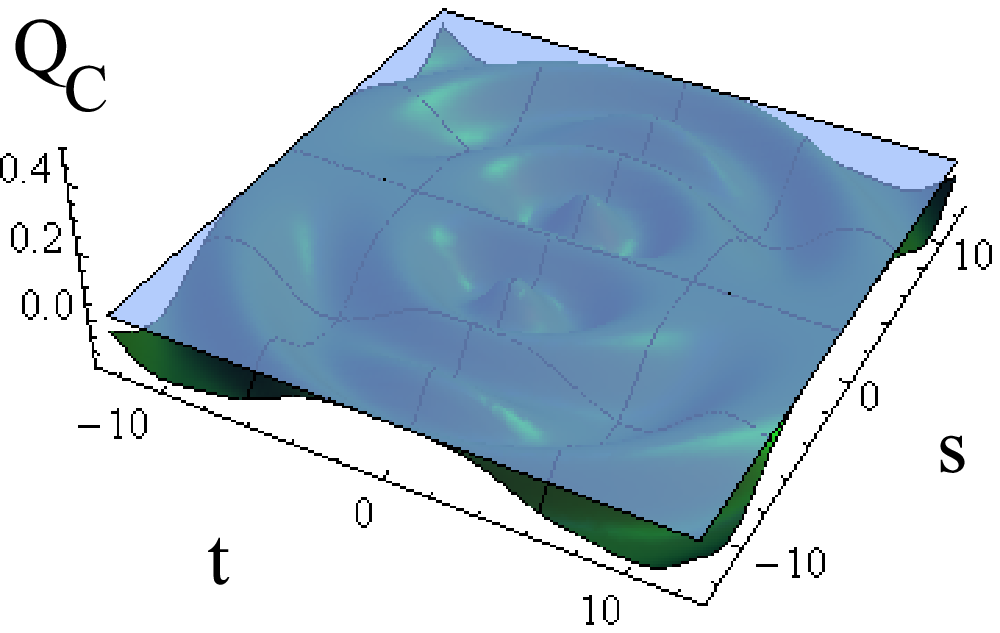}
\caption{\textbf{Uncorrelated system:} The heat flow pattern into $A,B,C$ for the ordinary product state $\rho_A \otimes \rho_B \otimes \rho_C$, with parameters $\lambda_A=0.15$, $\lambda_B=0.2 $, $\lambda_C=0.3$, $t$ for the interaction $H_{AB}$ and $s$ for $H_{BC}$. Here a light-coloured plane is shown through $Q=0$ to highlight those regions where heat leaves the subsystem $Q<0$, while the dark regions visible above this plane correspond to heat entering the system $Q>0$.}
\label{productABC}
\end{figure}

We now consider an interesting system consisting of 3 qubits $A$, $B$, and $C$ in a line, where $A$ is allowed to interact with $B$ and $B$ is allowed to interact with $C$. The state of the total system is given by a mixed state $\rho_{ABC}$, and has the property that $\rho_{AB}= \rho_A(T_A) \otimes \rho_B(T_B)$ and $\rho_{BC}= \rho_B(T_B) \otimes \rho_C(T_C)$ for thermal marginals $\rho_i(T_i)$, $i=A,B,C$ with $T_A <T_B<T_C$. Hence the pairs $AB$ and $BC$ obey the thermodynamic arrow individually and we would expect heat only to flow from $C$ to $B$ and from $B$ to $A$.

However, if $\rho_{AC}$ is given by
\begin{eqnarray}
\rho_{AC} &=&\frac{1}{2} ( ( \gamma + \lambda_C -\lambda_A) |10 \>\<10|+(\gamma -\lambda_C +\lambda_A) |01 \>\<01|\nonumber\\
&&\hspace{-1cm}+\sqrt{\gamma^2 - (\lambda_C -\lambda_A)^2}( |10 \>\<01|+|01 \>\<10|)\nonumber\\
&&\hspace{-1cm}+(  \lambda_A+\lambda_C- \gamma) |00 \>\<00|+ (2-\lambda_A -\lambda_C -\gamma) |11\>\<11|\nonumber\\
&&\hspace{-1cm}+\sqrt{(\lambda_A +\lambda_C -\gamma)(2-\lambda_A -\lambda_C-\gamma)}(|00\>\<11|+|11\>\<00|) )\nonumber
\end{eqnarray}
then we are able to exploit correlations between $A$ and $C$ to for example transfer heat from $A$ to $B$. The state $\rho_{AC}$ at first glance looks complicated, but it may be obtained from a 3 qubit state $|\Psi \>_{ACD}$ of the form (\ref{A1states}) by tracing out $D$; the parameters $\lambda_A, \lambda_C, \gamma$ must then obey similar relations to (\ref{constraints}).

The total state now takes the form $\rho_{ABC} = \rho_{AC} \otimes \rho_B$ and we assume that the interaction Hamiltonians are once again given by $V_{AB}= \frac{1}{2}(X_AY_B -Y_AX_B)$ and $V_{BC}= \frac{1}{2}(X_BY_C -Y_BX_C)$. In the event that only one of these interactions is switched on we have that the thermodynamic arrow is obeyed and heat always flows in the direction $C\rightarrow B \rightarrow A$. However when both interactions are on, the heat flow becomes more complicated.

If we assume the system $\rho_{ABC}$ evolves under the unitary $U(t,s) = \exp[-itV_{AB} - isV_{BC}]$, then Fig.~\ref{correlatedABC} and Fig.~\ref{productABC} show the contrast between heat flows for the entangled state $\rho_{ABC}$ and for the uncorrelated product state $\rho_A\otimes \rho_B \otimes \rho_C$.

\begin{figure}[t]
\centering
\includegraphics[width=4cm]{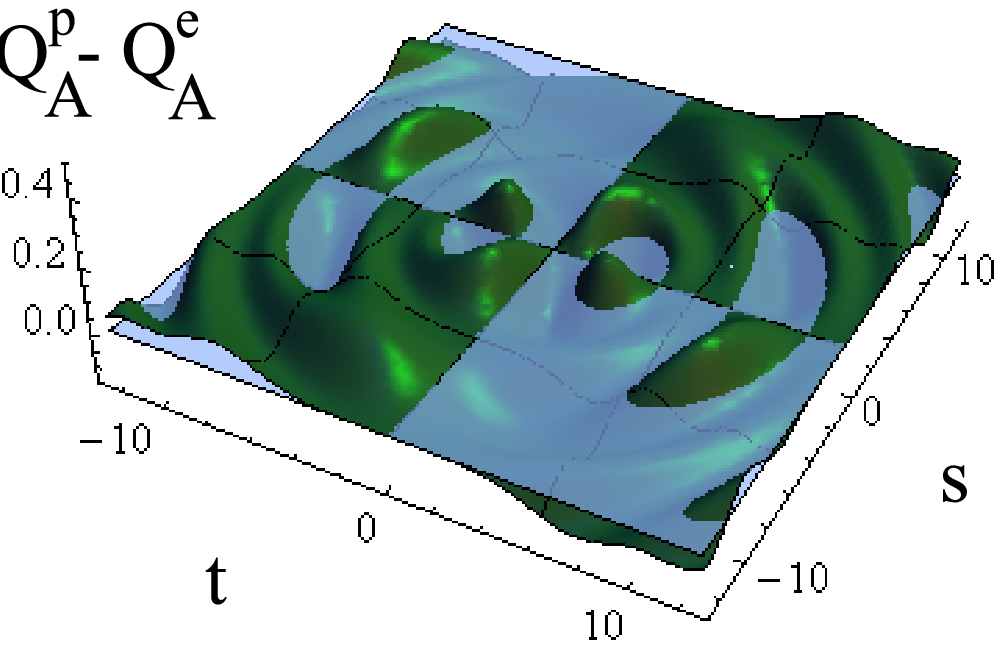}
\includegraphics[width=3cm]{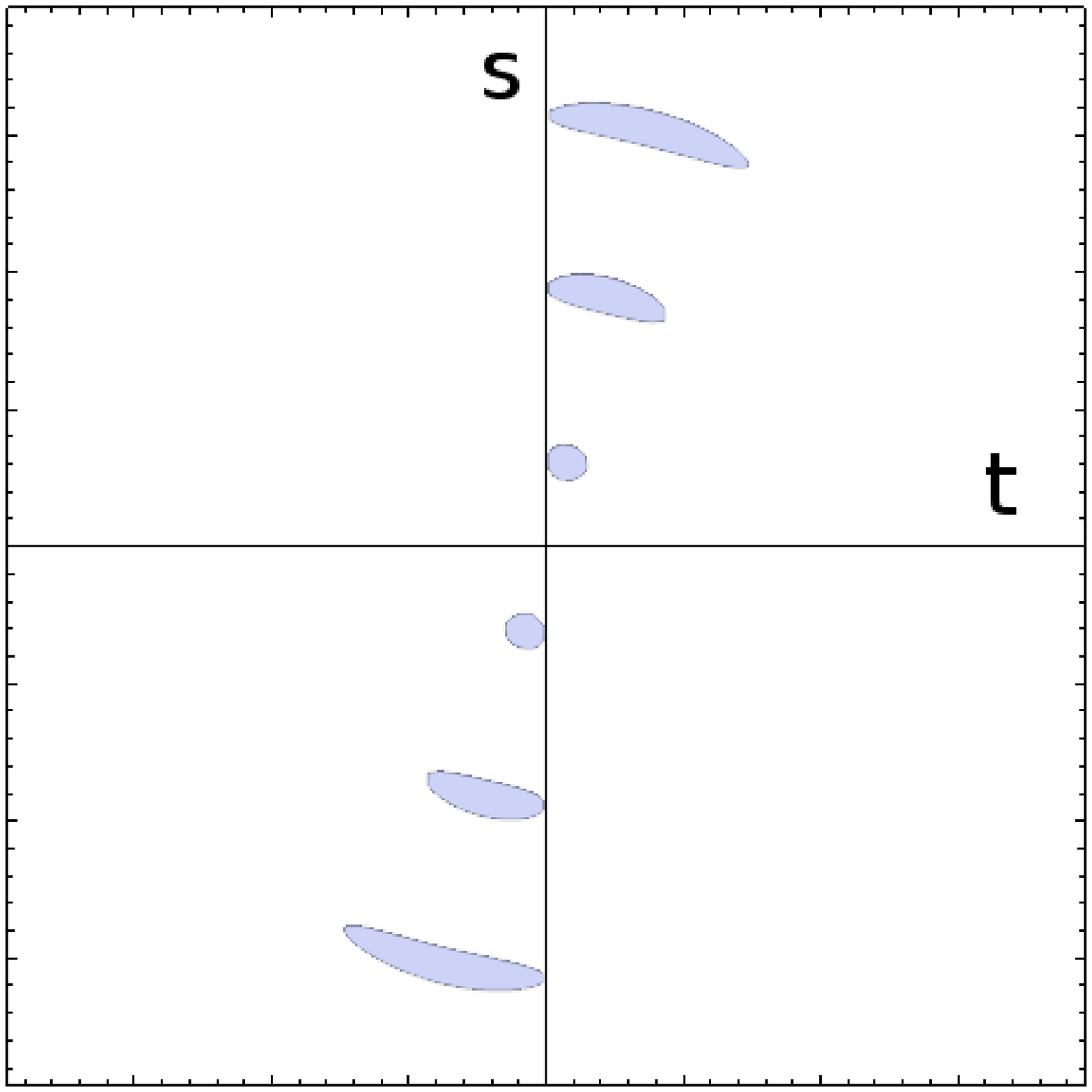}
\caption{The first graph shows the difference in heat pattern for $A$ in the two cases of an entangled state and a separable state. Here $Q^p_A$ is the heat gained by $A$ in the product state case, while $Q^e_A$ is the heat gained by $A$ in the entangled state case. Once again, light regions correspond to negative values, while dark regions correspond to positive values. The second plot shows the regions of $s$-$t$ parameter space in which $A$ has lost heat and both $B$ and $C$ have gained heat.}
\label{Delta-Q}
\end{figure}

For $s=0$ heat can only flow from $B$ to $A$, however as can be seen from Fig.~\ref{correlatedABC} by switching on $s$ we are able to have system $A$, the coldest subsystem, emit heat into $B$. Fig.~\ref{productABC} shows the expected heat flow patterns given the standard uncorrelated product thermal states. Here the heat flow into $A$ is always positive (dark-coloured regions), while the heat flow into $C$ is always negative (light-coloured regions). It is clear that the presence of entanglement can dramatically alter this pattern.
Fig.~\ref{Delta-Q} shows the difference $ \Delta Q_A \equiv Q_A^{p}-Q_A^{e}$ between the flow for $A$ in the two cases of a product state ($Q^p_A$) and an entangled state $Q^e_A$. It is evident from this that the entanglement can increase the heat \emph{into} $A$ as well as out of $A$. Fig.~\ref{Delta-Q} also shows the region of parameter space for which heat leaves $A$ and both $B$ and $C$ gain heat, which corresponds to a clear violation of the expected thermodynamic behaviour.

\begin{figure}[t]
\centering
\includegraphics[width=4.5cm]{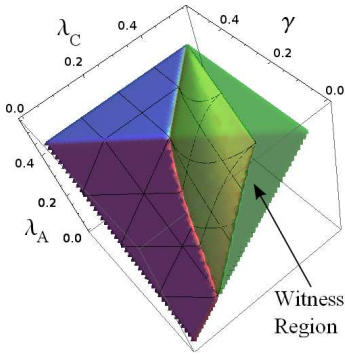}
\caption{The region of parameter space for which reversal of the thermodynamic arrow in $\rho_{ABC}$ can act as an entanglement witness is shown in semi-transparent light region, while the remainder of the space only possesses correlations that could be attributed to a classically correlated state.}
\label{witness-region}
\end{figure}

We can also ask if the system can act as an entanglement witness under heat flow. By analysing the mutual information for the system we can show that range exists in the parameters $(\lambda_A, \lambda_C,\gamma )$ for which the correlations present in the state $\rho_{ABC}$ exceed $\log 2$ and so allows heat flows that cannot be attributed to classical correlations between $A$ and $C$. Fig.~\ref{witness-region} shows this region as part of the diamond polygon parameter space discussed earlier. The set of states for which it may act as an entanglement witness is shown as the semi-transparent light-coloured region. However, it must be emphasized that for a generic, correlated mixed state it is not always possible to unitarily eliminate \emph{all} of the correlations present as the set of unitary orbits containing product states is a low dimensional sub-manifold of the manifold of all unitary orbits\cite{geometry-unitary-orbits}.

\subsection{A quiver of thermodynamical arrows}

As already mentioned in general there exists a multitude of global states $\rho_{\mbox{\tiny tot}}$ associated with a given combination of thermal marginal states. For a given collection of marginals the full state can have a vast array of different consistent correlations. The set of such consistent states $\Q$ contains states that differ in the correlations that exist between different combinations of subsystems. Consequently, given subsystems $S_1, \dots S_k$, the reduced state $\rho_{1 \cdots k}$ can vary for different global states $\rho_{\mbox{\tiny tot}}$ in $\Q$.

It is easy to see that while we might demand the individual subsystems $S_1, \dots, S_N$ to be thermal states, this thermality may appear at various levels of coarse graining on the collection of subsystems. For example we might have for a pair of subsystems $S_a$ and $S_b$ that $\rho_{ab}=\Tr[\rho_{\mbox{\tiny tot}}]= \rho_a \otimes \rho_b$, however for a triple of subsystems $S_a,S_b,S_c$ the state $\rho_{abc}$ is entangled. For such a situation, the thermodynamic arrow is present for any energy transfer process involving only the two subsystems $a$ and $b$, however for processes involving three or more subsystems the correlations present allow violations of the thermodynamic arrow.

Consequently, the different possible correlation structures that can occur in a state $\rho_{\mbox{\tiny tot}} \in \Q$ correspond to a hierarchy of thermodynamic arrows. The set of states $\Q$ can be partitioned up as $\Q= \A_1 \cup \A_2  \cup \cdots $, where a state $\rho$ is in $\A_k$ if there exists subsystems $S_{a_1}, \dots , S_{a_k}$ such that
\begin{eqnarray}
\sum^k _{j=1}  \frac{Q_j}{T_j} \ge 0
\end{eqnarray}
is observed for all transformations on $\rho_{a_1 \dots a_k}$ and no larger set of subsystems exist with this property. It is not clear, however, if a non-trivial partitioning can be defined in which, for a state in a given class, the thermodynamic arrow holds for \emph{all} marginals on $k$ subsystems.

This classification of states in terms of their largest thermal marginal depends on the particular correlation structure of the global state without reference to where the subsystem are situated, however in more practical situations it makes sense to include a notion of locality. For example, we might consider a collection of subsystems $S_1, S_2, \dots$ located at various points in space. For a given subsystem $S_a$ we may define the notion of an ``arrow range'', $R_a$, which is the largest $r_a$ such that the reduced state $\rho (r_a)$ on all subsystems within a distance $r_a$ of $S_a$ is a thermal marginal. For the entire state $\rho_{\mbox{\tiny tot}}$ we may then simply define a characteristic arrow range $\bar{R}$ as the average over the set $\{R_i \}$ of ranges for each subsystem $S_i$. States with large $\bar{R}$ do possess correlations, however these correlations are difficult to access in practice, requiring many coordinated, local pairwise interactions $V_{ij}$ in order to generate an effective multipartite interaction over the correlated state. On the other hand, states with small values for $\bar{R}$, for example states of the form (\ref{A1states}) or the state $\rho_{ABC}$ considered in the previous section, have correlations that are more easily accessible through local pairwise interactions.
This provides a simple generalization of the `local' Maxwell demon introduced in the first section. A demon should not only be finite in terms of its memory resources, but also in terms of the correlation range $R$ that it can probe.

\section{Conclusion}
Time, with its inexorable flow, is one of our oldest mysteries. In stark contrast, it is only in the last century that we have become aware of the quantum mechanical properties of Nature, and only in the past few decades that some of the subtleties and power of quantum correlations have revealed themselves.

In this article we have examined some aspects of entanglement correlations, how they can reverse the standard thermodynamic arrow of time, and have shown that in this regard thermodynamic transformations may act as entanglement witnesses.

We analysed how highly entangled multipartite states can make ordinarily forbidden thermodynamic states accessible. Different correlation structures can give rise to the same set of local thermal states, and so processes involving several subsystems are in general required to exploit the entanglement present. The set of transformations on a system then possess a hierarchy of thermodynamic arrows, whether considered globally or in terms of local correlation ranges. The key aspect is to allow interactions that activate the correlations present, which we demonstrated with a mixed state example $\rho_{ABC}$ in which the thermodynamic arrrow is in place for interactions involving only $A$ and $B$ or for $B$ and $C$, but by switching on an interaction with system $C$ this can be dramatically modified.

It would be of interest to further explore the multipartite correlation structures that can occur in a quantum state - for example we might speculate that in the early, dense universe that $\bar{R}$ is small enough to allow any random physical interactions to exploit the correlations present, producing a gradual disappearance of the thermodynamic arrow the closer we get to the initial state of the universe. Of course such notions of locality rely on a classical spacetime background, and so we could only push the issue of the special initial conditions into the sub-Planck scales, where a complete theory of quantum gravity would be required.

Another setting of interest is that of quantum field theory, where thermality arises naturally due to observers possessing causal horizons, such as in the case of black holes or for accelerated observers. Once again, thermality can be viewed as arising from entanglement across the horizon, however it is unclear if these correlations could actually be utilized in any thermodynamic transformations.  At a deeper level, it may also be fruitful to explore connections with the thermal time hypothesis due to Connes and Rovelli \cite{thermal-time} or recent work by Padmanabhan \cite{paddy} that attempts to relate the cosmological arrow of time for our expanding universe with the thermodynamic arrow by exploiting the thermodynamics of local spacetime horizons.

Finally, in a more grounded setting, it would be of interest to see if a mixed state $\rho_{ABC}$, or one like it, could be realised experimentally to see how easily one might exploit quantum correlations present in a state to affect its thermodynamic behaviour.

\end{document}